\documentclass[aps,prl,reprint,twocolumn,superscriptaddress]{revtex4-1} 
\usepackage{graphicx}
\usepackage{amsmath,amssymb}
\usepackage{multirow}

\newcommand{\bea}{\begin{eqnarray}}
\newcommand{\eea}{\end{eqnarray}}
\newcommand{\beq}{\begin{equation}}
\newcommand{\eeq}{\end{equation}}
\newcommand*\diff{\mathop{}\!\mathrm{d}}
\def\half{\frac{1}{2}}
\def\br{{\bf r}}
\def\dr{\diff\br}

\begin{document}
      
\title{A Simple Generalized Gradient Approximation for the Non-interacting 
Kinetic Energy Density Functional}
\author{K.~Luo}
\email{kluo@ufl.edu}
\affiliation{Quantum Theory Project, Department of Physics,
 University of Florida, Gainesville, FL 32611}
\author{V.V.~Karasiev}
\email{vkarasev@lle.rochester.edu}
\affiliation{Laboratory for Laser Energetics, University of Rochester,
 Rochester, NY 14623}
\author{S.B.~Trickey}
\email{trickey@qtp.ufl.edu}
\affiliation{Quantum Theory Project, Department of Physics and 
Department of Chemistry, University of Florida, Gainesville, FL 32611}

\date{22 June\ 2018}

\begin{abstract}
A simple, novel, non-empirical, constraint-based 
 orbital-free 
generalized gradient approximation (GGA) 
non-interacting kinetic
energy density functional is presented along with illustrative applications.  
The innovation is adaptation of constraint-based construction 
to the essential properties of pseudo-densities from the pseudo-potentials 
that are essential in plane-wave-basis {\it ab initio} molecular dynamics. 
This contrasts with constraining to the qualitatively 
different Kato-cusp-condition densities.   
The single parameter in the new functional is calibrated by satisfying Pauli potential positivity constraints for pseudo-atom densities.  
In static lattice tests on simple metals and semiconductors, the new LKT functional outperforms the 
previous best constraint-based GGA functional, VT84F (Phys.\ Rev.\ B \textbf{88}, 
161108(R) (2013)), is generally superior to a recently proposed meta-GGA, is reasonably competitive with   
parametrized two-point functionals, and is substantially faster.
\end{abstract}
\maketitle

{\it Introduction.} \noindent
Hohenberg-Kohn density functional theory (DFT)
\cite{HK64,Levy79}  has come to prominence mainly in  
Kohn-Sham (KS) orbital form \cite{KohnSham}.
 However, driving  {\it ab initio} molecular dynamics (AIMD)
\cite{Barnett93,MarxHutter2000,Tse2002,MarxHutter2009} with KS DFT exposes
a computational cost-scaling burden. The KS computational cost 
scales no better than $N_e^3$ with $N_e$ the number of
electrons or number of thermally occupied bands.
Additionally there is reciprocal space 
sampling cost or equivalent costs from large real-space unit cells
used with $\Gamma$-point sampling.  In contrast, orbital-free DFT
(OF-DFT) offers linear scaling with  
system size \cite{PostNAMETOFKEreview,WittEtAlReview} for use of AIMD on 
arbitrarily large systems.

The long-standing  barrier to
widespread use of OF-DFT has been the lack of reliable 
non-empirical approximate kinetic energy density functionals (KEDFs). 
In terms of the KS orbitals $\varphi_j$, the  exact, positive definite
KS kinetic energy (KE) density is \vspace*{-4pt}
\beq
t_s[n] = t_s^{orb} \equiv \half \sum_{j=1}^{N_e} \vert \nabla \varphi_j\vert^2 \,,
\label{eq:KStsubs}
 \vspace*{-4pt}
\eeq
in Hartree atomic units 
with $n(\br)$ the electron number density (and unit occupation for
simplicity). 
Two types of approximate KEDFs have been explored, semi-local (one-point)
 \vspace*{-4pt}
\beq
T_{s} [n] = \int\!\!\dr \, t_{s}[n(\br),\nabla n(\br), \ldots] \; 
\label{eq:OFtsubsdefn}  
\vspace*{-4pt}
\eeq
and two-point with a non-local term
 \vspace*{-4pt}
\beq
T_{NL}[n]\!\! =\! c_{TF}\!\!\! \iint\!\! \dr\! \dr' n^\alpha(\br)
K[n(\br),n(\br'),\br,\br'] n^\beta(\br')  
\label{eq:2point}
\vspace*{-4pt}
\eeq
with $c_{TF}= \frac{3}{10}(3\pi^2)^{\frac{2}{3}}$.
For a dimensionless $K$,  $\alpha + \beta= 8/3$. Most approximate 
$K$s are parametrized;  see Refs.\ 
[\onlinecite{WittEtAlReview,WangGovindCarter1999,HuangCarter2010,MGP2018,Constantin2018}]  for details as well as brief discussion below.   In this communication, we propose 
a novel non-empirical one-point KEDF and show that it is competitive with current two-point KEDFs, generally better than other one-point functionals, more transferable, and notably faster.

{\it Generalized Gradient Approximations.} \noindent The simplest one-point 
functionals are Thomas-Fermi \cite{Thomas27,Fermi27,Fermi28}
 \vspace*{-4pt}
\beq
T_{TF}[n] := \int\!\! \dr\, t_{TF}(\br) \; , \; %
t_{TF}(\br) : = c_{TF} n^{\frac{5}{3}}(\br) \;,
\label{eq:TF} 
 \vspace*{-4pt}
\eeq
and von-Weizs\"acker \cite{Weiz35}
 \vspace*{-4pt}
\beq
T_{W}[n] := %
\frac{1}{8}\int\!\! \dr \, \frac{|\nabla n(\br)|^2} {n(\br)} \equiv \int\!\! \dr\, t_{W}(\br) %
\,.
\label{eq:TwDefn}
 \vspace*{-4pt}
\eeq
Neither is satisfactory as a general KEDF.  
As with approximate exchange-correlation (XC) functionals \cite{Perdew92}, the 
gradient expansion of the weakly inhomogeneous electron
gas KE leads to consideration of generalized gradient approximations 
(GGA) for $T_s$,
 \vspace*{-4pt} 
\beq
T^{GGA}_{s}[n]=\int\!\! \dr \,t_{TF}(\br)  %
F_{t}(s(\br)) \, . 
\label{eq:A6} 
 \vspace*{-4pt}
\eeq
Here $F_t(s)$ is the GGA KE enhancement factor, a function of the dimensionless reduced density gradient $  s :=  \frac {|\nabla n|}{2nk_f} \equiv  \frac{1}{2(3\pi^2)^{1/3}}
  \frac{|\nabla n|}{n^{4/3}} $
familiar from GGA X functionals. GGA KEDFs so constructed automatically
satisfy $T_s$ uniform scaling requirements \cite{LevyPerdew85}.  In 
GGA form  the von Weizs\"acker KE becomes $ F_{W}(s) = \frac{5}{3}s^2$.

From the Pauli term decomposition \cite{PostNAMETOFKEreview,PRB80,Signpost}, 
 \vspace*{-4pt}
\beq
T_{s}[n]  = T_{W}[n]+T_{\theta}[n]  \; ,
\label{eq:Ttheta}
 \vspace*{-4pt}
\eeq
three constraints follow\cite{LevyOu-Yang88},
 \vspace*{-4pt}
\begin{eqnarray}
T_{\theta}[n] &\ge& 0 \label{eq:Tthetapositive} \,,\\
v_{\theta}(\br) &\ge& 0 \; \forall \; \br ,
\label{eq:vthetapositive}  \\
v_{\theta}(\br) & \geq & \frac{t_{\theta}(\br)}{n(\br)}\;\; \forall \; {\br},\quad t_{\theta}:=t_{s}^{orb} - t_{W}
\label{eq:paulipot_inequality}
\;,
 \vspace*{-4pt}
\end{eqnarray}
with the Pauli potential defined as $v_{\theta}(\br) := \delta T_{\theta}[n]/\delta n(\br)$ and the Pauli enhancement factor is  $F_{\theta}(s) = F_{t}(s) - F_{W}(s)$.

To date, perhaps the best constraint-based GGA KEDF 
is VT84F (evaluated at T=0 K of course)\cite{VT84F}.  It
is successful in finite-T AIMD simulations \cite{KCT2016}  and is the only 
non-empirical GGA KEDF that yields reasonable binding in simple solids.  
It was 
constrained to satisfy Eqs.\ (\ref{eq:Tthetapositive}) and 
(\ref{eq:vthetapositive}) for physical atom densities, i.e., those that obey 
the Kato cusp condition \cite{Kato57}.  
VT84F also was constrained to respect $\lim_{s\rightarrow \infty} F_{\theta}(s)/F_{W}(s)= 0 $.
This comes from the one-electron tail region of a many-electron atom\cite{LevyPerdewSahni84} where $t_\theta/t_{W}$  must 
vanish, hence $t_s \rightarrow t_{W}$ \cite{DreizlerGrossBook}. 

In terms of the universal Hohenberg-Kohn-Levy density functional, 
such a physically motivated constraint is non-universal:
the Kato cusp condition is specific to an external Coulomb potential.  Such
non-universality is rational for material and molecular property
calculations.  But the ubiquitous use of pseudo-potential plane-wave
basis methods in AIMD simulations means that it is not the optimal
non-universality for them.  OF-DFT calculations in fact require a
local pseudo-potential (LPP).  The OF-DFT Euler equation then implies that
$v_\theta$ is closely related to the LPP
$v_{\rm ext}^{\rm pseudo}$ and that $v_\theta$ is evaluated with the
corresponding pseudo-density.  Thus any constraint based on
density characteristics should be specific to a particular type or class of
pseudo-potential.

Ref.\ [\onlinecite{HarrisFestVVKSBT}] explored some elementary consequences 
for constraint satisfaction (or
violation) with non-Kato densities.
 Difficulties with simpler
one-point KEDFs (linear combinations of $T_{TF}$ and $T_{W}$) used 
with orbital-free projector augmented-wave pseudo-densities 
also have been reported 
\cite{LehtomakiEtAl2014}. So far as we know, no approximate KEDF has
been constructed by explicit satisfaction of the foregoing constraints,
Eqs.\ (\ref{eq:Tthetapositive})-(\ref{eq:vthetapositive}),
 for a specified type of pseudo-densities.   Nor has  
Eq.\ (\ref{eq:paulipot_inequality}) been used.  

\begin{figure}[]
  \includegraphics[width=0.95\linewidth]{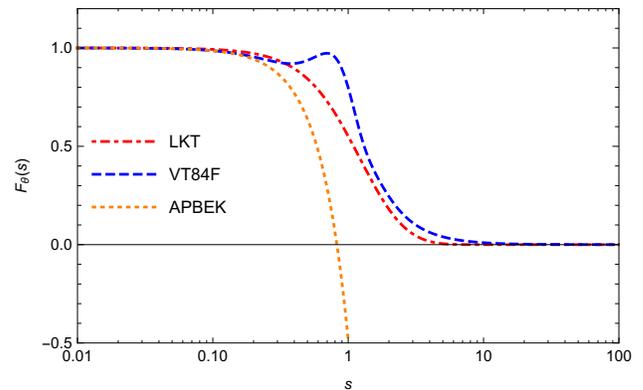}
  \caption{Pauli enhancement factors for LKT ($a=1.3$) (red dot-dashed), VT84F (blue dashed), and APBEK (orange dotted).}
  \label{fig:Ftheta}
\vspace*{-6pt}
\end{figure}

{\it New GGA KEDF.}\noindent We resolve this pseudopotential AIMD 
deficiency by devising a 
GGA KEDF constrained to satisfy  Eqs.\ (\ref{eq:Tthetapositive}) and 
(\ref{eq:vthetapositive}) for pseudo-densities of a particular kind
and show that in most spatial regions its $v_\theta$ 
satisfies Eq. (\ref{eq:paulipot_inequality}) as well.   
The new GGA KEDF enhancement factor is  
\beq
  F_{t}^{LKT}(s) = \frac{1}{\cosh(a\, s)} + \frac{5}{3}s^2 
  \label{eq:LKTGGA}
\eeq
with parameter $a > 0$.  Fig.~\ref{fig:Ftheta} compares $F_{\theta}^{LKT}$ with the 
VT84F and APBEK \cite{APBEK2011} enhancement factors.
It satisfies the obvious homogeneous electron gas constraint 
$\lim_{s \rightarrow 0} F_{t/\theta}(s) = 1$ 
and obeys  $0 \le F^{LKT}_{\theta} \le 1$ so as to 
 satisfy the bound conjectured by Lieb \cite{Lieb80,GazquezRobles82} 
\beq
  T_{s} \le T_{TF} + T_{W} \; .
  \label{eq:LiebConjecture}
\eeq
 $F_t^{LKT}$ also satisfies\cite{LevyOu-Yang88,LevyPerdewSahni84,Herring86} 
 $t_{\theta}([n];{\br})\; \ge 0 \; \forall\, {\br}$, thus $T^{LKT}_{\theta} \ge 0$. 

The sole parameter $a=1.3$ was determined as follows. 
A set of pseudo-densities was generated for the atoms H
through Ne with a typical Hamann norm-conserving non-local
pseudo-potential (NLPP) scheme \cite{HSC79} using default radii in the
APE code \cite{APEcode} and the Perdew-Zunger (PZ) XC local density
approximation (LDA) \cite{PZ81}.  Then $a$ was found such that all the
post-scf Pauli potentials from those pseudo-densities
 satisfied $v_\theta \ge 0 \; \forall \; r$.  Importantly, as long as an $a$ value gave
$v_\theta \ge 0$ for the H atom, positivity also was met for all the
heavier atoms.  For Li $a<1.4$ is required, while
for H, $a\le 1.3$ is needed to get a post-scf
$v_{\theta} \ge 0$. For He, the $a$ value does not seem to matter within the
range tested.  While the $a$ value is non-universal, we
expect reasonable transferability to those other 
pseudo-potential types for which the pseudo-densities are similar, 
specifically those with nearly flat
pseudo-densities near the nucleus.  The expectation is confirmed by post-scf 
and scf calculations for atoms.  

Though reference atom set, H--Ne, encompasses 
1--8 pseudo-electrons, equally good performance 
for other elements is not assured. Post-scf 
determination of $a$ also is distinct from self-consistent calculation, 
which might vitiate the supposedly constrained behavior. 
Atomic tests are the first line of investigating these issues. 
For a given pseudo-potential and XC approximation, self-consistent
solution of the KS equation provides the exact KS $t_\theta$ and the
ingredients to construct the exact KS Pauli $v_\theta$ (see Eq.~(35) in
Ref.\ [\onlinecite{PRB80}]). Those are the standards  
against which to judge $t_\theta$ and $v_\theta$ from an
approximate KEDF.  In anticipation of the OF-DFT calculations on 
periodically bounded systems reported below, we focused upon the bulk-derived
LPP (BLPS) \cite{BLPS2004,BLPS2008} for two atoms, 
Al and Li. Here we discuss Al because it was not in the $a$ calibration.
Li discussion is in the Supplemental Material \cite{SuppMat}. (The 
Li pseudo-atom is challenging because it is a one-orbital 
system ($2s^1$) for which $T_\theta$  should vanish.) 
 Again the XC functional is PZ. 

\begin{figure}[h]
	\centering
	\includegraphics[width=1.05\linewidth]{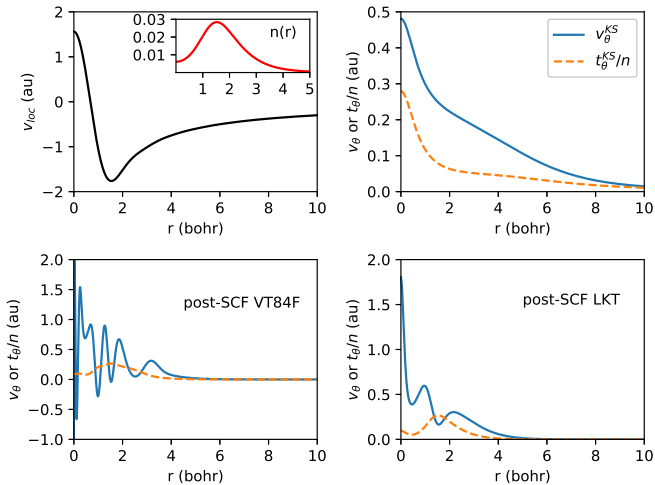}
	\caption[Al BLPS atom]{Upper left: Al BLPS as function of radial position (inset: KS pseudo density); 
Upper right: exact KS $v_\theta$ and $t_\theta/n$. Lower left: post-scf 
$v_\theta$ (solid) and $t_\theta/n$ for VT84F (dashed). Lower right: Same for LKT.}
	\label{fig:Al_ttheta_vs_vtheta_postscf}
\vspace*{-10pt}
\end{figure}

Fig.~\ref{fig:Al_ttheta_vs_vtheta_postscf} displays the exact
$t_\theta/n$ and $v_\theta$ for the BLPS Al pseudo-atom in the
 $3s^2\,3p^1$ configuration and the post-scf results with that
pseudo-density for both VT84F and LKT. 
Note several features.  Though VT84F was  
constructed 
to satisfy $v^{VT84F}_\theta \ge 0$ near a nucleus for Kato-cusped densities,
it also satisfies that constraint arbitrarily close to the nucleus for 
the cusp-less pseudo-density.  However, 
$v^{VT84F}_\theta$ becomes negative near $r=0.1$ bohr,   
 a clear example of the crucial non-universality.  
LKT does not have that problem.
Second, $v^{LKT}_\theta$ is much smoother 
than $v^{VT84F}_\theta$, though not as smooth as $v^{KS}_\theta$.  
Third, except for a small region around $r=1.8$ bohr, $v^{LKT}_\theta$
respects the Pauli potential inequality, Eq.\ (\ref{eq:paulipot_inequality}),
whereas  $v^{VT84F}_\theta$ violates it in four regions that span much
of the significant density magnitude. 

Note also that, unlike some other GGA KEDFs, e.g.\ 
 E00\cite{E00}, PBE2 \cite{Signpost}, and APBEK, $v_\theta^{LKT}(r)$ decays correctly to zero asymptotically for an atom.
This may be useful in the AIMD simulation of low-density regions 
of matter. Though $v_\theta^{VT84F}$ decays similarly, its rapid 
oscillations in the dominant density region might slow scf convergence rates
as well as cause other difficulties.

\begin{figure}[h]
	\centering
	\includegraphics[width=0.9\linewidth,]{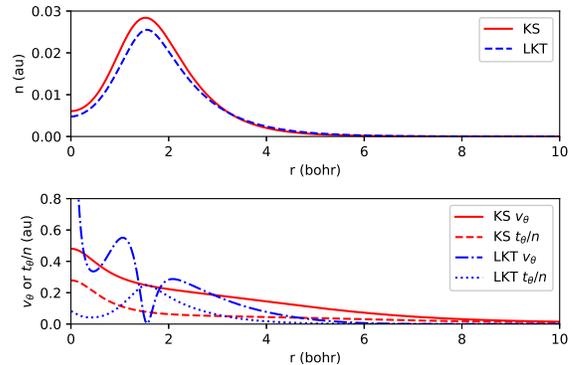}
	\caption[Al BLPS atom]{Top: Al KS (solid, red) and LKT(dashed, blue) pseudo-densities as function
of radial position. Bottom: KS vs.\ LKT $v_\theta$ (solid red vs.\ dash-dotted blue, upper pair) and similarly  $t_\theta/{n}$ (dashed red v. dotted blue; lower pair). 
}
	\label{fig:vthetaLKT}
\vspace*{-8pt}
\end{figure}

Self-consistent OF calculations for the BLPS Al pseudo-atom show that
$v_\theta^{LKT}$ stays positive, though it exhibits oscillations quite
similar to those seen in the post-scf case; see
Fig.~\ref{fig:vthetaLKT}. The inequality of
Eq.~(\ref{eq:paulipot_inequality}) is violated only around $r=1.8$
bohr as in the post-scf case. However, the LKT Pauli energy per particle 
is far from the KS value.

{\it Performance on Solids.} \noindent
Validation of the new functional for AIMD requires accuracy tests on 
extended systems. We therefore did KS-DFT and OF-DFT 
calculations on simple metals and semiconductors. Conventional 
KS calculations were done with {\sc Abinit} \cite{AbinitCode} and  the OF-DFT calculations used 
{\sc Profess}\cite{Hung..Carter10} and/or 
{\sc Profess}$@${\sc Quantum-Espresso} \cite{ProfessQE}.
Again the PZ LDA XC functional and BLPS were used. 
For comparison we included the Wang-Govind-Carter (WGC) \cite{WangGovindCarter1999}, Huang-Carter (HC) \cite{HuangCarter2010}, and Constantin et al. KGAP
\cite{Constantin2018} two-point KEDFs and the one-point Constantin et al.
SOF-CFD \cite{SOFCFDarxiv} meta-GGA (Laplacian-dependent) KEDF. Technical details 
and parameter values are in the Supplemental Material \cite{SuppMat}.

 Note that
WGC was parametrized for main-group metals and yields poor binding curves 
for semiconductors, while HC was
parametrized for semiconductors.  KGAP is
parametrized to experimental direct band gaps.  Results from the one-point
functionals E00, APBEK, and PBE2 are omitted because of unrealistic 
binding curves for the former two and instability problems for the latter one. 
KGAP comparisons are
from Tables I and II of Ref.\ [\onlinecite{Constantin2018}]. SOF-CFD values
are from Table I of Ref. [\onlinecite{SOFCFDarxiv}].  
Equilibrium volumes, energies, and bulk moduli for other functionals were 
generated by varying $\pm 5\%$ around the equilibrium volume to obtain
11 energy-volume points, which then were fitted to the 
Birch-Murnaghan equation of state \cite{BM47}.

The metals were Li, Mg, and Al in the simple cubic, body-centered
cubic, face-centered cubic, and hexagonal close-packed structures. Nine III-V semiconductors in zinc-blende structures
were treated: AlP, AlAs, AlSb, GaP, GaAs, GaSb, InP, InAs, and
InSb. 

\begin{table}[h]
	\caption{KEDF performance on solid metals and semiconductors: MARE of equilibrium volumes $V_0$, energies $E_0$, 
and bulk moduli $B_0$, as percentages. See text for notation. \\}
	\centering
	\begin{ruledtabular}
		\begin{tabular}{c|cccccc}			
			\multirow{2}{*}{} 
		KEDF	&	\multicolumn{3}{c}{Metals} & 	\multicolumn{3}{c}{Semiconductors}  
			\\
			\hline 
			&   $V_0$ & $E_0$ & $B_0$ &   $V_0$ & $E_0$ & $B_0$ \\
			WGC   &  0.7  & 0.0  &   2.7   &     -   &   -    &    -   \\
			HC    &  5.5  & 0.6  &  12.3   &    1.5  &  0.5   &   4.9  \\
			KGAP\footnotemark[1]   &  4.0  & -  &  5.1   &    3.0  &  -   &   16.2  \\

			VT84F &  6.0  & 0.1  &  11.6   &   10.5  &  3.6   &  56.4  \\
			SOF-CFD\footnotemark[1] &  5.2  & 0.6  &  8.5   &   3.4  &  0.9   &  10.0  \\
			LKT   &  4.0  & 0.2  &   7.7   &    2.1  &  2.8   &   4.3 
		\end{tabular}
	\footnotetext[1]{Note: only metals with cubic symmetry were included and PBE XC was used.} 
	\end{ruledtabular}
	
	\label{tab:MARE}
\end{table}

With KS quantities as references, Table ~\ref{tab:MARE} shows the mean
absolute relative error (MARE) percentages for equilibrium volume $V_0$, energy $E_0$ per atom (for metals) or per cell (for semiconductors), and bulk moduli $B_0$ from WGC,
HC, KGAP, VT84F, SOF-CFD, and LKT. These are calculated as  $|(Q_{\mathrm{OF}} -Q_{\mathrm{KS}})/Q_{\mathrm{KS}}|\times 100/N_{\rm systems}$, where $Q$ is $V_0$, $E_0$, or $B_0$. (More detailed tabulations are in the Supplemental Material \cite{SuppMat}.) For 
 $V_0$ and $B_0$, LKT is a significant
improvement over VT84F. The $V_0$ and $B_0$ MAREs are reduced by 33\% in 
metals.  The reduction is more dramatic in the semi-conductors, a factor of
5 for $V_0$ and 13 for $B_0$.  The semiconductor $E_0$ MARE is 
reduced by 22\% but worsened slightly from 0.1\% to 0.2\% for the
metals.   Except for performance on semi-conductor $E_0$, it also is
clear that the LKT GGA is superior to the more-complicated non-empirical 
SOF-CFD meta-GGA KEDF.  

Regarding the two-point functionals, WGC outperforms all the other
functionals on the metals but is inapplicable on semiconductors,
recall above \cite{WangGovindCarter1999}.  Conversely, HC with
averaged parameters exhibits balanced error, with all three MAREs
within $5\%$ (except $B_0$ for metals). KGAP does well on volumes in
both classes but not $B_0$.  Remarkably LKT exhibits performance
competitive with both HC and KGAP in prediction of equilibrium volumes
for both material classes. Moreover, LKT outperforms HC for $B_0$ and
is much more balanced than KGAP for $B_0$. (Comparison with the
recent MGP two-point functional is of no avail, since its
parametrization is tuned to match KS results for each system
\cite{MGP2018}.)

For the case of AlP, we found that LKT converges for
relatively smaller energy cutoff than needed with VT84F
and HC. Typically LKT also requires fewer 
self-consistent iterations for solution to a given tolerance 
than are needed by either HC or VT84F and each LKT iteration is 
typically about one-fifth the time of an HC iteration.  
Thus the one-point LKT is 
more useful as a broadly applicable 
functional than the highly parametrized two-point HC KEDF
or the experimentally parametrized two-point KGAP KEDF yet 
is simpler, faster, and mostly better than the SOF-CFD one-point KEDF. 
LKT seems therefore to be currently the most promising candidate for 
general AIMD OF-DFT use or with 
small-box algorithms \cite{Chen16}. Though it remains to be tested, we 
anticipate the finite-T generalization \cite{KarasievSjostromTrickey2012} 
of LKT will be of value for warm dense matter simulations.

As to limitations, LKT does not yield a good value of $V_0$ for bcc Li
with a 3-electron LPP. So far as we know, all GGA KEDFs 
developed so far share this limitation. The extent of transferability
to another distinct class of pseudo-potential, along with the post-scf
determination of $a$, remains to be examined. 

\begin{acknowledgments}
KL and SBT were supported  by U.S.\ Dept.\ of Energy 
	grant DE-SC 0002139.
VVK acknowledges support by the Dept.\ of Energy National Nuclear Security 
Administration under Award Number DE-NA0001944.
	
\end{acknowledgments}

\end{document}